\documentclass[aip,amsmath,amsthm,amssymb,citeautoscript,reprint]{revtex4-2}
\usepackage{graphicx, subfig, xcolor}
\usepackage{physics, siunitx}
\usepackage{enumitem}
\usepackage[ruled, linesnumbered]{algorithm2e}

\usepackage[todonotes={textsize=tiny}, final]{changes}

\usepackage{hyperref}
\hypersetup{
    pdfauthor = {Amartya Bose},
    pdftitle = {Variational Adaptive Gaussian Decomposition: Scalable Quadrature-Free Time-Sliced Thawed Gaussian Dynamics},
    colorlinks = true,
}

\SetKwComment{Comment}{// }{}

\begin{document}

\title{Variational Adaptive Gaussian Decomposition: Scalable Quadrature-Free Time-Sliced Thawed Gaussian Dynamics}
\author{Rahul Sharma}
\affiliation{Department of Chemical Sciences, Tata Institute of Fundamental Research, Mumbai 400005, India}
\author{Amartya Bose}
\email{amartya.bose@tifr.res.in}
\email{amartya.bose@gmail.com}
\affiliation{Department of Chemical Sciences, Tata Institute of Fundamental Research, Mumbai 400005, India}
\begin{abstract}
    Time-slicing has emerged as a strategy for incorporating semiclassical propagation into real-time path integral formulation and recovering full quantum dynamics. A central step is the decomposition of a time-evolved wave function into a superposition of Gaussian wave packets (GWPs). Here we introduce a quadrature-free variational framework for GWP decomposition, reformulating it as an optimization problem in which the GWP parameters are chosen to maximize the overlap with the time-evolving wave function. An autoencoder-decoder neural network is used for this optimization, with the representation being adaptively reoptimized during propagation. Each wave packet in this decomposition represents a localized patch of the underlying semiclassical manifold, while retaining full correlations between all degrees of freedom. This variational adaptive Gaussian decomposition (VAGD) approach yields a compact Gaussian expansion, providing a scalable route to time-sliced semiclassical quantum dynamics. While general, applying VAGD to facilitate time-slicing of thawed Gaussian dynamics allows a route to improving the semiclassical treatment to the full quantum mechanical result in a systematic manner.
\end{abstract}

\clearpage 
\onecolumngrid
\listofchanges[title={Major Revisions}, show=comment]
\clearpage 
\twocolumngrid

\maketitle

\section{Introduction}
The dynamics of molecular systems is governed by the time-dependent
Schr\"odinger equation. However, its direct numerical solution becomes
intractable as the number of degrees of freedom increases, as encountered in
grid-based approaches such as the split-operator Fourier transform (SOFT)
method~\cite{feitSolutionSchrodingerEquation1982,
kosloffFourierMethodSolution1983}, owing to the exponential growth of the
Hilbert space. Formally exact wavefunction propagation methods, including the
family of multiconfigurational time-dependent Hartree ((ML-)MCTDH)
approaches~\cite{meyerMulticonfigurationalTimedependentHartree1990,
mantheMultilayerMulticonfigurationalTimedependent2017}, extend the accessible
regime to systems with several tens of degrees of freedom, but remain ultimately
limited by the scaling of the underlying Hilbert space and specific forms of
potential. A different class of approaches that has emerged is the semiclassical
methods, which exploit classical structure to provide accurate and
computationally tractable approximations to quantum
dynamics~\cite{millerSemiclassicalMethodsChemical1986}.

Among these, two broad classes can be identified: initial value
representations based on stationary-phase approximations to the path
integral, such as the van Vleck–Gutzwiller propagator and the
Herman–Kluk formulation~\cite{vanvleckCorrespondencePrincipleStatistical1928,
gutzwillerPhaseIntegralApproximationMomentum1967,
hermanSemiclasicalJustificationUse1984,
millerAlternateDerivationHerman2002,
kayHermanKlukApproximation2006}, and Gaussian wavepacket (GWP)
dynamics, such as the thawed Gaussian approximation (TGA)~\cite{hellerTimedependentApproachSemiclassical1975,
hellerTimeDependentVariational1976,
hellerSemiclassicalWayMolecular1981,
hellerCellularDynamicsNew1991,
vanicekFamilyGaussianWavepacket2023}. In TGA, a single Gaussian
wavepacket is propagated with time-dependent center, width, and
phase, enabling efficient on-the-fly \textit{ab initio}
simulations~\cite{begusicOntheflyInitioSemiclassical2018,
begusicOntheflyInitioThree2018,
begusicSingleHessianThawedGaussian2019,
begusicApplicabilityThawedGaussian2022,
begusicFiniteTemperatureAnharmonicityDuschinsky2021}. However, the
accuracy of such local semiclassical approximations typically
deteriorates at long times due to the breakdown of the underlying
harmonic expansion.

A natural route to improved accuracy is to represent the wavefunction
as a superposition of Gaussian wavepackets. This idea underlies a wide
range of approaches, including variational multiconfigurational
Gaussian methods (vMCG) and Gaussian-based MCTDH
formulations~\cite{burghardtApproachesApproximateTreatment1999,
  maioloMultilayerGaussianbasedMulticonfiguration2021}, which
determine Gaussian parameters through the Dirac-Frenkel time-dependent variational principle and have
been successfully applied to system–bath
problems~\cite{burghardtMulticonfigurationalSystembathDynamics2003,
  burghardtMultimodeQuantumDynamics2008a,
  romerGaussianbasedMulticonfigurationTimedependent2013}. Adaptive
basis-growth strategies, such as multiple spawning methods, introduce
new Gaussians dynamically to capture non-adiabatic
effects~\cite{ben-nunInitioMultipleSpawning2000,
  ben-nunMultipleSpawningApproach2000}, while trajectory-guided
Gaussian bases and related approaches employ classical or
semiclassical trajectories to construct efficient
representations~\cite{sallerBasisSetGeneration2015}.
Matching-pursuit-based decompositions combined with SOFT propagation
(MP/SOFT) further exploit Gaussian expansions to obtain efficient
representations of quantum dynamics of system-bath decomposed
problems~\cite{wuMatchingpursuitSimulationsQuantum2003,
  wuQuantumTunnelingDynamics2004,
  wuMatchingpursuitsplitoperatorFouriertransformSimulations2005}. Related
approaches also include coupled coherent Gaussian (CCG) methods ~\cite{shalashilinMultidimensionalQuantumPropagation2001,
shalashilinPhaseSpaceCCS2004} and
trajectory-driven Gaussian basis techniques, which propagate Gaussian
ensembles along classical or Ehrenfest-like
trajectories~\cite{sallerBasisSetGeneration2015} or guided by quantum
trajectories~\cite{guQuantumDynamicsGaussian2016,
  dutraQuantumDynamicsQuantum2020,
  dutraMultidimensionalTunnelingDynamics2020}. More recently, Gaussian
dynamics has also been formulated variationally using Rothe’s method
of
propagation~\cite{rotheZweidimensionaleParabolischeRandwertaufgaben1930,
  schraderTimeEvolutionOptimization2024,
  schraderMultidimensionalQuantumDynamics2025}. While these approaches
differ in their exact formulation, they share a common objective of
constructing flexible, potentially time-dependent Gaussian bases
capable of capturing quantum dynamics at a manageable cost.

An alternative strategy for improving long-time accuracy is
time-slicing~\cite{hellerWavepacketPathIntegral1975,
burantRealTimePath2002,
kongTimeSlicedThawedGaussian2016}, in which the propagation is divided
into short intervals over which semiclassical approximations remain
valid. At the end of each interval, the wavefunction is re-expressed as
a superposition of Gaussian wavepackets, which serve as initial
conditions for subsequent propagation. In the limit of sufficiently
fine slicing, this procedure converges to exact quantum dynamics.
However, practical implementations face two key bottlenecks: (i) the
need for high-dimensional quadrature to perform Gaussian decomposition,
and (ii) the rapid growth in the number of Gaussian components required
for accurate representation. The former is particularly severe for
oscillatory quantum wavefunctions, leading to the well-known Monte
Carlo sign problem~\cite{makri1988MonteCarloPathIntegration,makri1987MonteCarloOscillatoryIntegrands,mak1990SolvingSignProblem}, which worsens dynamically as complex phases develop
during time evolution. The second affects performance by increasing the number
of classical trajectories that require simulation.

Time-sliced thawed Gaussian dynamics (TSTG) has demonstrated that
highly accurate results can be obtained using quadrature-based
Gaussian
wavepacket decompositions with each GWP being propagated using TGA~\cite{kongTimeSlicedThawedGaussian2016}. However, the
exponential scaling of the number of Gaussian components with
dimensionality remains a fundamental limitation for high-dimensional
and \textit{ab initio} applications. Unlike TSTG, related approaches, such as
the basis expansion leaping multiconfiguration Gaussian (BEL-MCG)
method~\cite{kochBasisExpansionLeaping2013,
murakamiAccurateQuantumMolecular2018, murakamiNonadiabaticQuantumMolecular2019},
combine periodic re-expansion with evolving coefficients, but do not evolve the
Gaussian basis over time. The semiclassical evolution of the basis is often an
efficient way of increasing the length of the segments in between the
re-expansion steps without loss of accuracy.

In this work, we introduce a variational adaptive Gaussian
decomposition (VAGD) framework that addresses these limitations by
reformulating the Gaussian decomposition step in time-sliced dynamics as a
global optimization problem. Specifically, the parameters of the output Gaussian
ensemble are determined variationally to maximize the overlap with the input
wavefunction using a minimal number of wavepackets resulting without losing accuracy. This replaces the conventional quadrature-based
decomposition with a nonlinear optimization over the space of Gaussian
parameters, effectively converting time-sliced semiclassical dynamics into a
quadrature-free, optimization based framework. The resulting problem is highly
non-convex, but can be efficiently handled in practice using an
autoencoder–decoder neural network that maps an input Gaussian ensemble to a
compact reparametrized representation. The systematic route of going from
semiclassical to fully quantum mechanical results provided by a combination of
VAGD-enabled time-sliced dynamics makes it attractive for complex
multidimensional problems.

This paper is organized as follows. Time-sliced thawed Gaussian dynamics is
reviewed in Sec.~\ref{sec:tstga}. The VAGD framework is introduced in
Sec.~\ref{sec:vagd}, followed by numerical demonstrations in
Sec.~\ref{sec:numerics}. We conclude in Sec.~\ref{sec:conclusion}.

\section{Methods}
\subsection{Time-Sliced Thawed Gaussian Dynamics}\label{sec:tstga}
The time-evolution of an initial wave-function $\ket{\psi_0}$ of a $d$-dimensional system, under a
Hamiltonian,
$\hat{H} = \frac{1}{2}\sum_j\frac{\hat{P}_j^2}{M_j} +
V\left(\left\{\hat{X}_j\right\}\right)$ is given by $\ket{\psi(t)} =
\exp\left(-\frac{i}{\hbar}\hat{H}t\right)\ket{\psi_0}$. Depending upon the
particular semiclassical approximation used, the propagator,
$\exp\left(-\frac{i}{\hbar}\hat{H}t\right)$, is represented in different ways
using classical trajectories. Under the thawed Gaussian approximation
(TGA)~\cite{hellerTimedependentApproachSemiclassical1975}, the time-dependent
wave function is presumed to be approximated as a GWP:
\begin{align}
  \braket{\vec{x}}{\psi(t)} &= \exp\left(\frac{i}{2\hbar}\left(\vec{x} - \vec{q}_t\right)^\intercal A_t \left(\vec{x} - \vec{q}_t\right)\right.\nonumber\\
  &+ \left.\frac{i}{\hbar}\vec{p}_t^{\,\intercal}\left(\vec{x} - \vec{q}_t\right) + \frac{i}{\hbar}\gamma_t\right)\label{eq:gwp}
\end{align}
where under the local harmonic approximation, the real ``position'' and
``momenta'' coordinates, $\vec{q}_t$ and $\vec{p}_t$ respectively,
satisfy the classical equations of motion,
\begin{align}
  \dot{\vec{q}}_t &= M^{-1}\vec{p}_t\label{eq:tga_pos}\\
  \dot{\vec{p}}_t &= -\vec{\nabla} V(\left\{\vec{q}_t\right\})\label{eq:tga_mom}
\end{align}
where $M$ is the diagonal mass matrix. The ``width'' matrix, $A_t$, and the
generalized phase, $\gamma_t$ satisfy the following equations:
\begin{align}
  \dot{A}_t &= -A_t M^{-1} A_t - \laplacian V(\left\{\vec{q}_t\right\})\label{eq:tga_A}\\
  \dot{\gamma}_t &= \frac{i \hbar}{2} \Tr(M^{-1}A_t) + \frac{1}{2}\vec{p}_t^{\,\intercal} M^{-1} \vec{p}_t - V(\left\{\vec{q}_t\right\}).\label{eq:tga_gamma}
\end{align}

While exact for harmonic systems, owing to the impact of the non-zero
higher-order derivatives of the anharmonic potentials, the quality of
traditional TGA deteriorates with time even when the initial state is a
Gaussian. Several improvements have been done on single GWP dynamics over the
years motivated by using time-dependent variational
principle~\cite{hellerTimeDependentVariational1976,
faouPoissonIntegratorGaussian2006, vanicekFamilyGaussianWavepacket2023,
moghaddasifereidaniHighorderGeometricIntegrators2023} and also local expansions
beyond the harmonic
approximation~\cite{moghaddasifereidaniHighorderGeometricIntegrators2024}.
Though these serve to increase the accuracy of the single GWP dynamics for
anharmonic systems at greater computational costs, all of them are still
approximations and cannot guarantee formal or numerical exactness in general.
Even the improved accuracy of these methods eventually fails although at longer
times. As outlined originally in
Ref.~\citenum{hellerWavepacketPathIntegral1975}, this deterioration can be
alleviated by recasting the semiclassical propagator in a time-sliced
form~\cite{burantRealTimePath2002, kongTimeSlicedThawedGaussian2016}, which can
be viewed as a discrete time approximation to the full quantum time-evolution
operator. The ``short-time'' propagators are approximated semiclassically. 

The essential idea is to divide the total propagation time into short time
segments, over which the local harmonic approximation is valid. Within each
segment, TGA, or any other semiclassical method, provides an accurate
approximation to the dynamics. At the end of these segments, the resulting wave
function is re-expressed as a superposition of Gaussians, which then serve as
initial conditions for the semiclassical propagation in the next time segment.
Thus, through an interleaving of the semiclassical propagation and decomposition
into GWPs, one can approximate the fully quantum dynamics at increasing levels
of accuracy. Ultimately, in the limit of infinitesimally short propagation
segments between successive decompositions ($\tau_\text{seg} =
N_\text{seg}\Delta t$), this time-sliced semiclassical dynamics method
approaches exact quantum dynamics.

Methods such as vMCG also represent the wavefunction as a superposition of
Gaussian wave packets, with both coefficients and basis parameters evolving
continuously in time~\cite{
burghardtApproachesApproximateTreatment1999,
burghardtMulticonfigurationalSystembathDynamics2003,
burghardtMultimodeQuantumDynamics2008a,
romerGaussianbasedMulticonfigurationTimedependent2013,
maioloMultilayerGaussianbasedMulticonfiguration2021}. In contrast, time-sliced semiclassical dynamics provides
an alternative framework in which the individual wavepackets are propagated over
short time segments using semiclassical dynamics, followed by a reconstruction
step at the segment boundaries, where the wavefunction is re-expressed as a
potentially different superposition of Gaussians, resulting in a simultaneously
updated set of coefficients and a redefined Gaussian basis for the subsequent
propagation. \added{Within each propagation segment, the equations of motion,
Eqs.\,\eqref{eq:tga_pos}\,--\,\eqref{eq:tga_gamma}, do not change the expansion
coefficients.} The time dependence within the segment enters through the
independent thawed Gaussian propagation of the Gaussian parameters.
\added{However, because the Gaussian basis is nonorthogonal, the overlap matrix
between different GWPs changes continuously during propagation. Consequently,
even though no coupled equations of motion are solved for the expansion
coefficients within a segment, the norm of the resulting superposition is
generally not conserved under the independent propagation of the individual
GWPs. A normalization step is therefore applied before the evaluation of
observables. This is equivalent to a common time-dependent rescaling of all
coefficients, while leaving their relative amplitudes and phases unchanged.
Avoiding explicit coefficient equations of motion preserves the simplicity of
the propagation scheme and allows the use of well-established thawed-Gaussian
trajectory algorithms.}\comment{Correction to coefficient evolution.} Thus,
unlike vMCG-type formulations, we do not solve coupled equations of motion for
the expansion coefficients within a segment. This corresponds to a piecewise
approximation to the full quantum propagator. The accuracy of this approximation
is controlled by the duration of the propagation segments and the quality of the
reconstruction step.

One possible implementation of this re-expansion can be in terms of the
over-complete coherent state basis. However, an immediate challenge that ensues
is that the resultant multidimensional integrals are not amenable to Monte Carlo
simulations owing to the accumulation of pure phase with time-propagation. A
Husimi transformed expansion-based implementation of this time-sliced thawed Gaussian has
recently been demonstrated~\cite{kongTimeSlicedThawedGaussian2016}. Despite its
success, the Husimi-based TSTG remains quadrature-dependent and scales poorly
with system dimensionality, motivating the need for a quadrature-free,
variational approach to GWP decomposition.

\subsection{Variational Adaptive Gaussian Decomposition (VAGD)}\label{sec:vagd}
We introduce a variational approach for constructing an optimized Gaussian
representation of an arbitrary input wave function. In the context of VAGD with
some Gaussian wavepacket dynamics, the normalized input wave function is written
as a sum of $N_\text{inp}$ Gaussians, $\ket{\psi} = \sum_{j=1}^{N_\text{inp}}
c_j\ket{\phi_j}$. The goal is to find a new representation in terms of
$N_\text{out}$ Gaussians, $\ket{\Psi} = \sum_{j=1}^{N_\text{out}}
C_j\ket{\Phi_j}$. Here, $N_\text{out}$ is capped at a certain user-defined
maximum number, $K$. An exact decomposition requires that the magnitude of the
overlap of the normalized input and output wave functions, called the fidelity, $F =
\frac{\abs{\braket{\psi}{\Psi}}}
{\sqrt{\braket{\psi}{\psi}\braket{\Psi}{\Psi}}}$, be unity. We, therefore, develop a variational
approach to maximizing this fidelity to a threshold value, $F_\text{thresh}$.
The exact $N_\text{out}$ that is used is adjusted adaptively, with the
\emph{optimal} representation being defined as the one that achieves the target
fidelity using a minimum $N_\text{out}$.

\begin{figure}
    \centering
    \includegraphics[width=\linewidth]{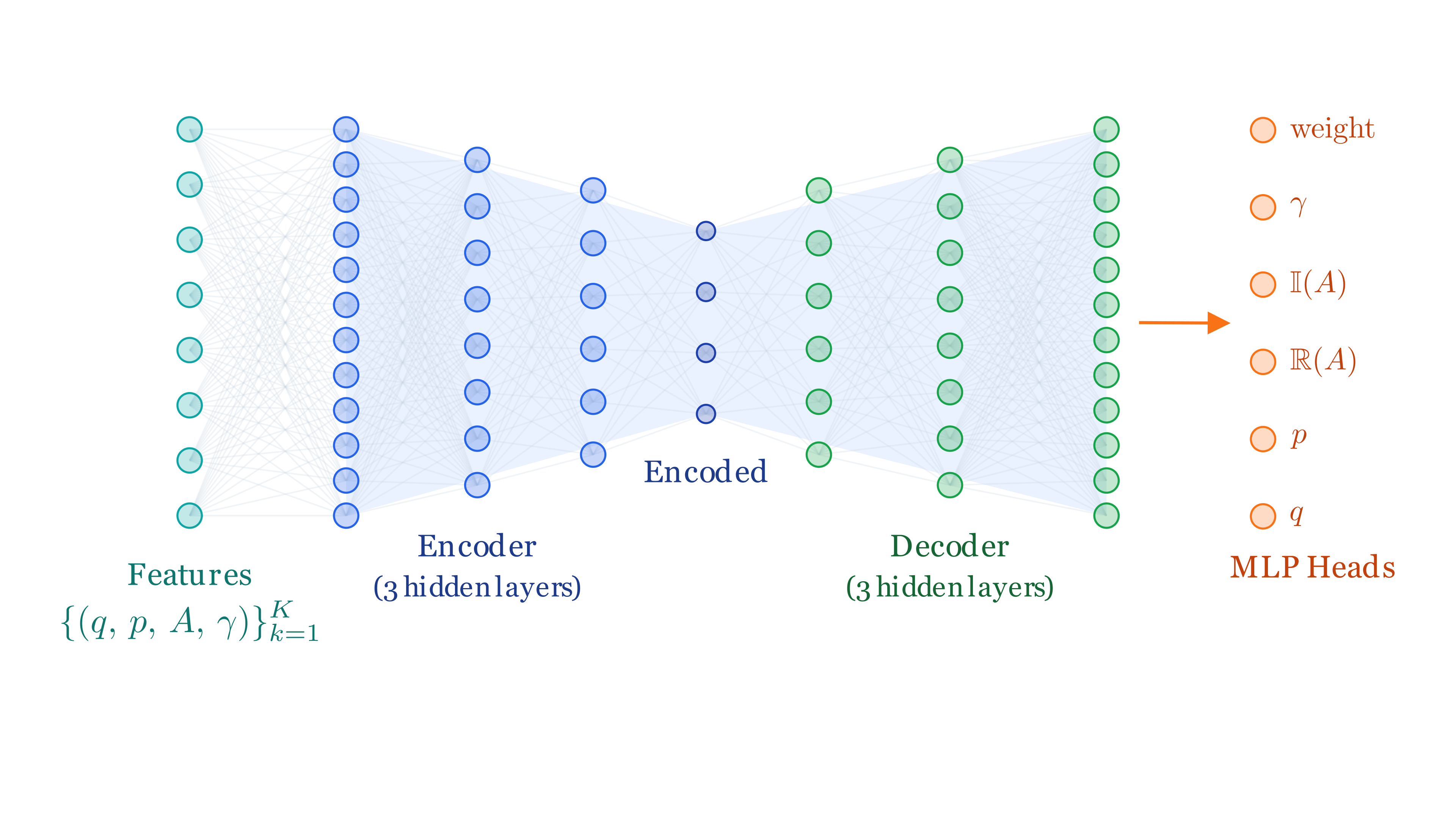}
    \caption{Schematic of autoencoder-decoder setup which acts as the basis transformation.}
    \label{fig:aed-basis-transform}
\end{figure}

The optimized representation is obtained through a variational procedure that
minimizes a loss defined in terms of the fidelity,
\begin{align}
    \mathcal{L} &= -\log(F) + (1-F),\label{eq:loss}
\end{align}
which is equivalent to maximizing the fidelity between an input wave function
and its Gaussian expansion.

The particular form of the loss has been chosen to accelerate the optimization.
The $-\log(F)$ term provides a sharp gradient when $F$ is small leading to
faster optimization, leading to faster convergence in the early stages of
optimization when the initial representation may be far from optimal. At these
stages, the $1-F$ term does contribute significantly to the gradient of the
loss. On the other hand, for the larger values of $F$ obtained towards the end
of the process, both the $-\log(F)$ and the $1-F$ terms contribute substantially
to the gradient, yielding faster convergence. The optimization is implemented
using an autoencoder-decoder neural network. Alternate approaches to
optimization are possible, but were found to be significantly less efficient and
stable in practice owing to the highly non-convex nature of the problem ~\cite{bottouOptimizationMethods2018}. (The
optimization procedure requires several evaluations of the fidelity, which is
obtained as a result of multiple multi-dimensional integrals. However, because
the wave functions are written as a superposition of GWPs, well-known
closed-form analytical results can be derived, which are summarized in
Appendix~\ref{sec:fidelity}.) Methods like MP/SOFT
~\cite{wuMatchingpursuitSimulationsQuantum2003, wuQuantumTunnelingDynamics2004,
wuMatchingpursuitsplitoperatorFouriertransformSimulations2005} also perform
similar optimizations in a sequential manner. However, VAGD leverages an
autoencoder-decoder network to provide a general framework for performing a
global optimization of the basis.

Figure~\ref{fig:aed-basis-transform} shows a schematic of the architecture,
which maps the input wave function to the set of parameters defining the output
representation via a low-dimensional ``encoded'' layer. The network is optimized
such that the loss, which depends upon the input and the output representations,
is minimized, ensuring that the resulting parameters encode a faithful Gaussian
decomposition of the input wave function.

It is important to emphasize that, unlike typical uses of
autoencoder-decoder architectures~\cite{bengioRepresentationLearningReview2013,Goodfellow-et-al-2016} in which the low-dimensional latent
``encoded'' space representation
\cite{hintonReducingDimensionalityData2006} is of interest, here the focus is on
the optimized output of the decoder. This output, $\ket{\Psi}$,
encodes a different but faithful representation of the input wave
function, $\ket{\psi}$. In this framework, the neural network serves
merely as a numerical optimizer for obtaining the parameters for the
Gaussian decomposition. With this perspective, it is not necessary to
train a generically applicable neural network. Instead, the network is
optimized specifically for each wave function under consideration.
A key point of this decomposition is that it is only dependent on the wave
function and not on the underlying potential energy surface.  Additionally,
because the overlap and the fidelity is constructed on a GWP-basis, we can get
analytical expressions for them, making the entire procedure quadrature free.
(The expressions are given in Appendix~\ref{sec:fidelity}.)

An optimal representation is defined by the minimum number of Gaussians used
($N_\text{out}$) to achieve $F_\text{thresh}$, essential for reducing the number
of classical trajectories. However, as end-users, we only set $K$, which is an
upper limit to $N_\text{out}$. The algorithm starts with an initial guess of
$N_\text{out}=N_\text{start}$. If upon optimization, the threshold fidelity,
$F_\text{thresh}$, is not achieved, then the value of $N_\text{out}$ is
increased by a user-specified step size $\Delta N$ and the network is
reoptimized. The process is repeated till either the threshold fidelity is
successfully achieved or the upper limit of $K$ is reached. We refer to this
procedure as an adaptive basis expansion technique, which ensures a near optimal
Gaussian decomposition. (Some important aspects of the encoding of the GWPs and
the techniques employed for efficiency and numerical stability are outlined in
Secs.~\ref{sec:encoding} and~\ref{sec:stabilization} respectively. Full details
of the particular version of the algorithm used in Sec.~\ref{sec:numerics} are
provided in the supporting information.)

\subsubsection{Encoding of Parameters}\label{sec:encoding}
To construct the input to the neural network and reconstruct the wave function
from the neural network, it is necessary to specify the parameters. A generic
$d$-dimensional GWP, Eq.~\eqref{eq:gwp}, is parameterized by two $d$-dimensional
real ``phase space'' vectors, $(\vec{q}, \vec{p})$, the $d\times d$ complex $A$
matrix, and the single complex number, $\gamma$, representing the phase.
Considering that the $A$ matrix has to be symmetric, $\frac{d(d+1)}{2}$ complex
numbers, can uniquely encode it. Therefore, $d^2+3d+2$ real numbers can be used
to represent a single GWP. So the input to the autoencoder-decoder has
$N_\text{inp} \times (d^2 + 3d + 2)$ input parameters.

Now the interpretation of the output is largely similar, with the $(\vec{q},
\vec{p})$  and  $\gamma$, retained. The main difference comes in the $A$ matrix:
the real part of $A$, $A^\text{R}$, must be symmetric, but the imaginary part of
$A$, $A^\text{I}$, which encodes the wave packet width, must additionally be
positive definite. Directly optimizing $A^\text{I}$ during training does not
guarantee this property. To ensure positive definiteness by construction, we
optimize a lower triangular matrix, $L$, with positive diagonal elements for
each GWP. The width matrix of the $k$th Gaussian is finally reconstructed using
a Cholesky decomposition~\cite{golubMatrixComputations2013}, $A_k^\text{I} = L_k L_k^T$.

The input layer to the encoder, therefore, encodes the input wave
function, and the output layer of the decoder encodes the output wave function.
Because the autoencoder-decoder network optimizes the output in a manner to
minimize the loss between the input and output~\cite{Goodfellow-et-al-2016,bengioRepresentationLearningReview2013}, in our case, we effectively get
a new representation of the input wave function. The output of the encoder part
of the neural network is an abstract low-dimensional latent space description of
the wave function. The general principle works irrespective of the particulars
of the neural network, and consequently, the network itself can also be treated
as a ``convergence'' object.

\added{For more difficult decompositions, the autoencoder-decoder neural-network
output can also be used as the starting point for an additional first-order
refinement. In this step, the decoded GWP parameters are optimized directly with
respect to the same fidelity-based loss, while retaining the same
parametrization and constraints. Thus, positive definiteness of the width matrix
and any imposed width restrictions are preserved by construction. This optional
second step of direct optimization provides a local refinement of the neural
network output without altering the adaptive structure of the
algorithm.}\comment{Optional second ``direct optimization'' step for more
difficult problems.}

This completes the discussion on the basic version of the VAGD algorithm.  There
are some numerical considerations that help with optimizing the network faster.
We discuss these subtleties in the next section.

\subsubsection{Numerical Stabilization Strategies}\label{sec:stabilization}

VAGD, as defined so far, allows for arbitrarily broad Gaussians to be
incorporated in the decomposition if required. While broad Gaussians may help
minimize the number of trajectories, the local harmonic approximation is
violated sooner for them than for the narrow wave packets, necessitating more
frequent decompositions to maintain accuracy. This issue can be mitigated by
putting a dimension-specific, user-defined cap on the width of each Gaussian,
which effectively sets a lower limit on the diagonal values of imaginary parts
of $A_k$, such that $(A^\text{I}_k)_{j,j}\ge (W_\text{min})_j$ for all
dimension, $j$.

A direct implementation of a VAGD decomposition step as described till
now starts from a randomly initialized network, which increases the
number of optimization cycles, or epochs, required for convergence to
$F_\text{thresh}$. Additionally, the position and the width of the
wave function can fluctuate significantly during time-evolution, which
makes the numerical optimization of the neural network difficult. We
will see that these two difficulties are closely related.

A natural way to address the randomly initialized starting network problem is to
use a ``warm-start'' ansatz. The fundamental intuition is that between
consecutive decomposition steps, the wave function may not have significantly
changed in shape. Instead of starting from a random network, one can therefore
reuse the optimized network from the previous step, potentially reducing the
number of epochs of training required.

However, while the intuition holds for the shape of the wave function, its
position can still change quite substantially between decomposition steps. As a
result, a direct warm-start ansatz may still lead to unstable or ineffective
optimizations. To stabilize the numerics and also maximize the effectiveness of
the warm-start ansatz, the wave functions are ``regularized'' to some common
scale.

This regularization consists of two preprocessing steps. First, the wave
function is recentered to remove its average position. Second, the coordinate
axes are rescaled so that the wave function has variances comparable to the
initial state. The neural network is trained on this preprocessed wave function.
After optimization, the regularization transformation is undone to recover the
physical representation.

\begin{algorithm*}
  \caption{The VAGD-TG Algorithm}\label{alg:vagd-tga}

  \KwData{$\ket{\psi(0)}$, $N_{\mathrm{start}}$, $N_{\mathrm{seg}}$, $K$, $F_{\mathrm{thresh}}$, $t_{\mathrm{final}}$, $\Delta t$, $\Delta N$}

  $n\gets 0$\Comment*[l]{Decomposition number}
  $t_\text{prev}\gets 0$\Comment*[l]{Segment begin time}
  $t_\text{next}\gets 0$\Comment*[l]{Segment end time}
  $N_\text{traj}^n\gets 0\,\forall n$\Comment*[l]{Number of Gaussians required at the $n$th decomposition}
  \added{$\ket{\psi_{\mathrm{in}}^{(0)}} 
    \xleftarrow{\mathrm{normalize}} \ket{\psi(0)}$}
    \Comment*[r]{All fidelities, diagnostics, and observables use normalized wave functions}
  \While{$t_\mathrm{prev}<t_{\mathrm{final}}$}{
    \eIf{$n=0$}{
      $N_{\mathrm{out}}\gets N_{\mathrm{start}}$\;
    }{
      \eIf{$\mathrm{jump}==\mathrm{false}$}{
        $N_{\mathrm{out}} \gets N^{(n-1)}_\text{traj}$\Comment*[l]{No jump, so use previous step's number}
      }{
        Take care of the possibility that the jump was a numerical optimization artifact:
        $N_{\mathrm{out}} \gets N^{(n-2)}_\text{traj}$\Comment*[l]{Previous step had a jump. Use the number of GWPs in the $n-2$th step}
      }
    }

    $\ket{\psi_{\mathrm{in}}^{(n)}} \xrightarrow[\;N_{\mathrm{out}}\;]{\mathrm{VAGD}}
    \ket{\Psi_N^{(n)}}=
    \sum_{j=1}^{N_{\mathrm{out}}}c_j^{(n)}\ket{\phi_j^{(n)}}$;\quad Compute the reconstruction fidelity $F_n$; \quad Warm start if $n>0$\;

    \eIf{$F_n<F_\mathrm{thresh}$ \textrm{and} $n\ge 2$} {
      jump $\gets \text{true}$\; }{
      jump $\gets \text{false}$\;
    }
    \While{$F_n<F_{\mathrm{thresh}}$ \textrm{and} $N_{\mathrm{out}}<K$}{
      $N_{\mathrm{out}}\gets\min\!\left(N_{\mathrm{out}}+\Delta N,K\right)$\;

      $\ket{\psi_{\mathrm{in}}^{(n)}} \xrightarrow[N_{\mathrm{out}}]{\mathrm{VAGD}}
      \ket{\Psi_N^{(n)}}=
      \sum_{j=1}^{N_{\mathrm{out}}}c_j^{(n)}\ket{\phi_j^{(n)}}$;\quad Compute the reconstruction fidelity $F_n$\;
    }

    Accept $\ket{\Psi_{\mathrm{VAGD}}^{(n)}}\gets\ket{\Psi_N^{(n)}}$ as the compact basis, $N^{(n)}_\text{traj}\gets N_\text{out}$\;

    $t_\text{next}\gets\min\!\left(t_{\mathrm{final}},t_\text{prev}+N_{\mathrm{seg}}\Delta t\right)$\;

    TGA propagation of each accepted $\ket{\phi_j^{(n)}}$ on $[t_\text{prev},t_\text{next}]$ \added{and normalization is done while calculating observables}\;

    \added{$\ket{\psi_{\mathrm{in}}^{(n+1)}} \xleftarrow{\mathrm{Normalize}} \ket{\psi_{\mathrm{out}}^{(n)}}$, $n\gets n+1$, $t_\text{prev}\gets t_\text{next}$\;}
  }
    \end{algorithm*}
\section{VAGD as a route to time-sliced thawed Gaussian dynamics}
VAGD allows an accurate representation of a given wave function as a
superposition of adaptive, optimally chosen GWPs. Many Gaussian wave
packet-based methods require such a description. Here we present the details of
using VAGD to enable time-slicing of thawed Gaussian
dynamics~\cite{kongTimeSlicedThawedGaussian2016}, a combination that we call
VAGD-TG. The idea is to propagate an initial wave function, $\ket{\psi(0)}$,
under a known Hamiltonian, $H$.

In addition to the time-step for the classical trajectories ($\Delta t$) used in the thawed Gaussian simulations, a VAGD-TG simulation is also characterized by
\begin{itemize}[nosep, topsep=0.5em]
    \item the number of time-steps between two decomposition steps ($N_\text{seg}$)
    \item the maximum number of Gaussians allowed for any of the decomposition step ($K$)
    \item the maximum allowed width of the individual GWPs making up the superposition. This is specified by the $W_\text{min}$ parameter which provides a lower bound on the imaginary part of the width matrix.
    \item The $F_\text{thresh}$, which is the minimum fidelity considered acceptable.
\end{itemize}

VAGD is applied to the $\ket{\psi(0)}$ to represent it as a sum of
$N_1$ GWPs, $\ket{\psi(0)} = \sum_{j=1}^{N_1}\ket{\phi^{(1)}_j}$. Even
if $\ket{\psi(0)}$ is itself a GWP, it might get broken down into
several, especially if its width is larger than the threshold. For
this step of decomposition, the autoencoder-decoder network has to be
started from random values, because there is no reference to be
leveraged for a warm-start. For each of the $N_1$ GWPs obtained from
the 1st decomposition, $\phi^{(1)}_j$, separate and independent thawed Gaussian
trajectories are launched following
Eqs.~\eqref{eq:tga_pos}--\eqref{eq:tga_gamma} for $N_\text{seg}$ time-steps.

Now for the $j$th decomposition ($j\ge 2$), we collect and regularize
all the propagated GWPs, and utilize the network from the $j-1$th step as a
warm-start ansatz as described in Sec.~\ref{sec:stabilization}. If the required
fidelity cannot be reached with $N_{j-1}$ GWPs, then it is increased in
user-defined steps of $\Delta N$ till the threshold fidelity is crossed upon
fitting. (Currently, we rerun the entire VAGD optimization using $N+\Delta N$
GWPs. However, one can easily think of doing a matching
pursuit~\cite{wuMatchingpursuitSimulationsQuantum2003} on top of the initial
VAGD result to incrementally improve it.)

If the previous decomposition step ($j-1$) required an increase in the number of
GWPs from the $j-2$th step, then we start with $N_{j-2}$ to ensure that jump was
not an optimization artifact. From the resulting decomposition, we start the
independent thawed Gaussian simulations for $N_\text{seg}$ time-steps. Depending
on the system under study, one can implement several very simple improvements.
One possible avenue which has not been incorporated for simplicity is to also
have an avenue to decrease the number of GWPs in the representation. In case the
target fidelity is reached immediately, one could try lower number of GWPs. This
would help in minimizing the number of trajectories. These interleaved thawed
Gaussian propagation and VAGD decomposition steps continue till the final time
of simulation is reached. (The full algorithm is summarized in Algorithm
\ref{alg:vagd-tga}.)

The accuracy of the resultant dynamics obtained from the thawed Gaussian
simulations is high in short time-scales, and for thinner GWPs because the local
harmonic condition is more valid across their length-scales. Therefore, by
decreasing the maximum width of the GWPs and shortening the length of the trajectory
segments, we can improve the accuracy of the semiclassical propagation for each
GWP. However, if the wavepackets are thin, then a larger number of them would be
required to describe the same wave function adequately, increasing the number of
trajectories. Therefore, in the limit of $K\to\infty$, $W_\text{min}\to \infty$
(corresponding to arbitrarily thin Gaussians), and the time between two
decompositions going to zero (or an infinite frequency of decomposition),
VAGD-TG is formally guaranteed to converge to the true quantum result. Obviously
this increases the cost of computation, and the goal is to get to this limit at
the minimum computational complexity. The variational optimization ensures that,
for fixed $K$, the representation is optimal in the sense of maximizing
fidelity.

Three factors go into considerations of the cost of a time-sliced thawed
Gaussian dynamics simulation: (1) the cost of individual classical trajectories
for a system, (2) the number of classical trajectories required, and (3) the
cost and number of decomposition steps required. As long as the system is the
same, the cost of the individual classical trajectories should remain constant
across different methods of simulation. Then the cost of a simulation is
governed by the number of trajectories and the cost of the decomposition. As we
shall demonstrate in Sec.~\ref{sec:numerics}, VAGD requires significantly (often
orders of magnitude) fewer trajectories than the previous implementations of the
time-sliced idea~\cite{burantRealTimePath2002, kongTimeSlicedThawedGaussian2016}
by leveraging a basis of GWPs with arbitrary phases, and arbitrary widths within
a maximal bound. Resolutions into coherent state basis typically use GWPs of
constant width and phase as the bases functions. The matching pursuit algorithm
initially used such coherent state
bases~\cite{wuMatchingpursuitSimulationsQuantum2003,
wuQuantumTunnelingDynamics2004,
wuMatchingpursuitsplitoperatorFouriertransformSimulations2005}. The last
consideration is one of the cost of every decomposition. In case of VAGD, this
would be governed by the complexity of the neural network involved. For simple
cases, the cost of the neural network becomes a significant proportion of the
simulation, but for more complicated systems especially using \textit{ab initio}
forces, the neural network cost, being independent of the force and energy
calculations, becomes less important. Further work aimed towards optimizing 
the autoencoder-decoder setup is ongoing and would be a big help in reducing the
cost of this step. (A simple empirical demonstration of the cost of our current
implementation is shown in Appendix~\ref{sec:cost}.)

\section{Numerical Illustrations}\label{sec:numerics}
We demonstrate the utility of VAGD in the context of VAGD-TG simulations on a
variety of anharmonic systems and deep tunneling systems. The split-operator
Fourier transform (SOFT)~\cite{feitSolutionSchrodingerEquation1982,
kosloffFourierMethodSolution1983} method is used to generate the numerically
exact dynamics that will be the point of comparison for the numerical examples
shown here. However, in the absence of exact reference data, VAGD-TG also
provides internal convergence criteria. At each decomposition step, the fidelity
threshold $F_\text{thresh}$ ensures that the wavefunction is accurately
represented within the chosen Gaussian basis, thereby controlling the projection
error. Convergence of the full dynamics can then be assessed by systematically
increasing the control parameters $K$ and $W_\text{min}$, and decreasing $N_\text{seg}$, and
monitoring the stability of observables of interest. Once these observables
become insensitive to further increases in computational effort, convergence is
established. This two-tiered approach, fidelity-based control of the
representation and observable-based convergence of the dynamics, provides a
practical and systematic route to convergence in the absence of exact solutions.

The specific details for the particular neural network used in the examples are
provided in the supporting information.

\subsection{Morse Potential}

We start our exploration with the dynamics of wave packets in Morse potentials.
In Sec.~\ref{sec:1Dmorse}, we explore the convergence of VAGD-TG to the full
quantum results under different levels of anharmonicity. Subsequently in
Sec.~\ref{sec:multiMorse}, we study the scaling of the algorithm by looking at
the dynamics under multidimensional independent Morse oscillators.

\subsubsection{1D Explorations}\label{sec:1Dmorse}

\begin{figure}
  \centering
  \includegraphics{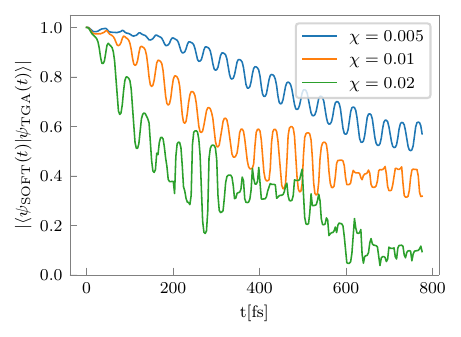}
  \caption{Breakdown of TGA with anharmonicity of 1D Morse potential.}
  \label{fig:tga-1dmorse-breakdown}
\end{figure}

First consider a 1D Morse oscillator:
\begin{align}
  V(x) &= D_e [1 - \exp(-a (x - x_\text{eq}))]^2,
  \label{eq:morse-1d}
\end{align}
where $D_e$ is the dissociation energy, $a$ is the width of the well, and
$x_\text{eq}$ is the equilibrium bond distance. It is often more convenient to
think in terms of the harmonic frequency at the bottom of the well, $\omega =
a\sqrt{2D_e}$, and an anharmonicity factor, $\chi =
\frac{\omega}{4D_e}$\cite{begusicSingleHessianThawedGaussian2019}, which is zero
for harmonic potentials. The parameters, $D_e$ and $a$ can be simply obtained by
inverting the relations. 

In Fig.~\ref{fig:tga-1dmorse-breakdown}, we show the overlap of the wave function evolved using the traditional TGA with the numerically exact
SOFT simulations for different values of $\chi$. Following the examples in
Ref.~\citenum{begusicSingleHessianThawedGaussian2019}, the equilibrium bond
length $x_\text{eq} = \SI{20.95}{\bohr}$ and $\omega=\SI{900}{\per\cm}$. The
initial wave packet is taken to be a real Gaussian centered at $q_0=0$, with $A
= \frac{i \omega_0}{\hbar}$, where $\omega_0 = \SI{1000}{\per\cm}$. The
simulations were run with a time-step of $\Delta t = \SI{0.053}{\fs}.$. The
reference dynamics for this one-dimensional model was obtained from a converged
SOFT calculation on a uniform grid over $x\in[-200,200]$, with grid spacing
$\Delta x=0.025\unit{a.u.}$. Consistent with our expectations, the quality of
TGA deteriorates with time, and this deterioration is faster for larger values
of $\chi$. 

\begin{figure}
  \centering
  \includegraphics{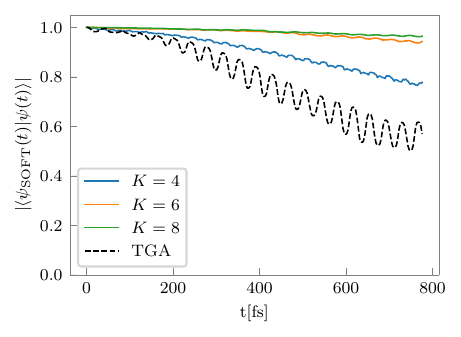}
  \caption{Overlap of the VAGD-TG method with the SOFT wave function for different maximum number of Gaussians.}
  \label{fig:tga-chi1-overlap-vs-nclusters}
\end{figure}

\begin{figure}
    \centering
    \includegraphics{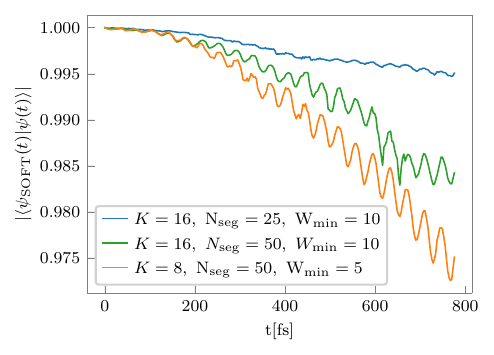}
    \caption{Variation of accuracy with respect to width of clusters and segment lengths.}
    \label{fig:width-nseg}
\end{figure}

Now, we explore our VAGD method. We perform the variational decomposition with
$N_\text{seg} = 50$. At each decomposition boundary, the reconstructed state is
required to satisfy $F_\text{thresh} = 0.9995$ with the propagated wavefunction.
The width parameter, $W_\text{min}$, was set to 5. 

Consider the case of $\chi=0.005$ as a first example. In
Fig.~\ref{fig:tga-chi1-overlap-vs-nclusters}, we examine the improvement of VAGD
with increasing $K$. There is a clear and systematic improvement of the overlap
with the SOFT result with the maximum number of GWPs allowed, $K$. This
improvement is most clearly observed at longer times, as the quality of the
semiclassical approximation gets worse. By $K\ge 6$, the VAGD-TG becomes
practically identical to SOFT. For this example, we push the limits and claim
convergence at $K=8$. \added{Additionally, we demonstrate the effect of
decreasing the length of the semiclassical segments and for thinner constituent
GWPs in Fig.~\ref{fig:width-nseg} and how these parameters help achieve improved
agreement with the exact quantum results in a systematic manner. On increasing
the $W_\text{min}$ parameter from 5 to 10, it was seen that just 8 GWPs cannot
adequately describe the wave function over all times. Thus it becomes necessary
to increase $K$. Then we see an improvement as a combination of these two
effects. On decreasing the $N_\text{seg}$ from 50 to 25, we see a further
improvement of the overlap.}\comment{Effect of change of $N_\text{seg}$ and
$W_\text{min}$. Convergence to full quantum result.} We explore how the VAGD
algorithm places the GWPs at different time-points for this problem in
Appendix~\ref{sec:nature-gwp}.

\begin{figure}
  \centering
  \includegraphics{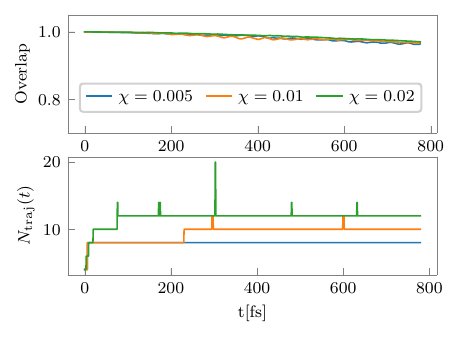}
  \caption{Overlap of VAGD-TG wave function with SOFT solution and number of VAGD trajectories of 1D-Morse oscillator for different values of $\chi$.}
  \label{fig:overlap-1dmorse}
\end{figure}

On increasing anharmonicity, while $\chi=0.01$ converged with the same
$N_\text{seg}$, $\chi=0.02$ required more frequent decompositions with
$N_\text{seg}=25$. We show the overlap and the number of trajectories for the
converged simulation in Fig.~\ref{fig:overlap-1dmorse}. The convergence gets
more difficult with anharmonicity. One way to think about this is greater the
anharmonicity, quicker is the breakdown of the local harmonic approximation for
GWPs of the same width. Therefore, we need to perform the VAGD steps more
frequently.

The number of trajectories, $N_\text{traj}(t)$, also shown in
Fig.~\ref{fig:overlap-1dmorse}, grows with time demonstrating the
build-up of the effects of anharmonicity and deviation from the basic
thawed Gaussian ansatz. However, it is noteworthy that the number of
trajectories required is relatively small in all the cases. This
becomes especially important when running \textit{ab initio} dynamics
where trading the cost of a better decomposition for a smaller number of
trajectories can prove to be extremely lucrative. For some of the runs, there
are spikes in the $N_\text{traj}(t)$ values. These sudden jumps come from the
optimizer getting stuck in local minima of the loss function. Step 14 in
Algorithm~\ref{alg:vagd-tga} takes care of this possibility and uses the number
of steps from the penultimate step to try the fitting. Moreover, depending on
the physics of the problem at hand, the actual number of trajectories may be
substantially less than the upper limit $K$ provided. This is another benefit of
the adaptive algorithm. We emphasize that checking the overlap with the
numerically exact SOFT results is by far the most stringent test of the
correctness of the method. It requires the approximate state to reproduce not
only the spatial distribution but also the detailed phase structure of the exact
quantum wave function.  Therefore, even a moderate reduction in the overlap does
not necessarily imply a comparable loss in the accuracy of physically relevant
observables, that would converge with far fewer trajectories. \added{This is
demonstrated numerically in the supporting information for the current case
where the autocorrelation spectrum for the $\chi=0.005$ system is already
converged at $K=6$ despite the overlap not being tightly
converged.}\comment{Convergence of spectrum vs overlap provided in SI}

\subsubsection{Multidimensional Explorations}\label{sec:multiMorse}

\begin{figure}
    \centering
    \includegraphics{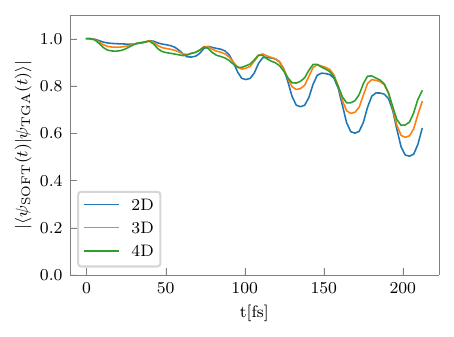}
    \caption{Overlap of TGA solutions with the SOFT solution for multidimensional Morse models as a function of dimensionality.}
    \label{fig:multi-morse-TGA}
\end{figure}
Pushing the semiclassical simulations to the full quantum levels of accuracy
gets increasingly difficult with dimensionality. The time-slicing procedure, if
done using quadrature, can very quickly become intractable. To study the scaling
of VAGD with dimensions, consider the simple case of $d$ uncoupled Morse
oscillators as an example:
\begin{equation}
    V(\vec{x}) = D \sum_{i=1}^{d}\left(1-\exp\!\left[-a\left(x_i-x_{0,i}\right)\right]\right)^2,
\end{equation}
where all the centers at origin ($x_{0,i}=0.0$) and $\omega =
\SI{900}{\per\cm}$ and $\chi = 0.005$ for all dimensions. While the scaling with
dimensionality would of course be worse for cases where the different dimensions
are correlated, this independent dimensional model allows us to study the
scaling in a relatively simplified setting. The VAGD-TG simulation does not
``know'' that the dimensions can be separated and hence does not use that
information explicitly. We will explore correlated multidimensional models in
terms of the double-well potentials. 

For the multidimensional Morse calculations, the initial wavefunction is taken as
\begin{equation}
  \Psi(\mathbf{x},0) = N \exp\left[-\frac{\omega}{2\hbar}
  \sum_{i=1}^{d}\left(x_i + \sqrt{\frac{8}{d}}\right)^2\right],
\end{equation}
with zero initial momentum. We start by demonstrating the
deterioration of the quality of the standard TGA simulation for
different dimensionality of the system in
Fig.~\ref{fig:multi-morse-TGA}. A time-step of
$\Delta t = \SI{0.053}{\fs}$ was used for all the
calculations. Because the dimensions are independent and the
displacement of the initial wave packet along each dimension decreases
with $d$, the errors in the single GWP TGA is less pronounced for
larger $d$.

\begin{figure}
    \centering
    \includegraphics{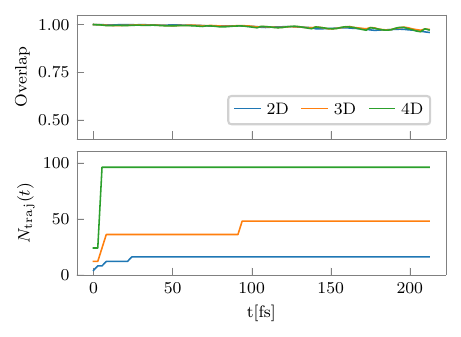}
    \caption{Overlap of VAGD-TG wave function with SOFT solution and number of trajectories for multidimensional Morse models as a function of dimensionality.}
    \label{fig:multi-morse-overlaps}
\end{figure}

Next, we perform the VAGD procedure every 25 steps. At each decomposition
boundary, the VAGD-reconstructed state is required to satisfy $F_\text{thresh} =
0.9995$ with the propagated and the $W_\text{min}$ for individual Gaussian wave
packets was set to 5.0 along each dimension. The results shown in
Fig.~\ref{fig:multi-morse-overlaps} demonstrate the recovery of full quantum
mechanical results. In all these cases, the number of trajectories saturated to
the maximum limit allowed, $K$. However, there are a couple of things to be
noted. First, the number of trajectories does not immediately converge to $K$.
Depending upon the exact system setup, it can use less trajectories as well.
This is the adaptive nature of the algorithm. The threshold $K$ that is required
is also a function of the duration of simulation. The longer a wave packet
propagates in an anharmonic potential, the more non-Gaussian character it
develops, and the higher is the number of trajectories that are required.

\begin{figure}
    \centering
    \includegraphics{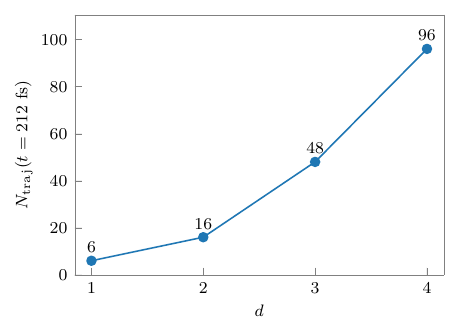}
    \caption{Number of trajectories required for converged results at $t=\SI{212}{\fs}$ for different numbers of dimensions.}
    \label{fig:scale_graph}
\end{figure}

Finally, in Fig.~\ref{fig:scale_graph}, we summarize the number of trajectories
required to obtain the converged results at a final time of
$t_\text{final}=\SI{212}{\fs}$. It is important, however, to interpret
Fig.~\ref{fig:scale_graph} only as an empirical summary of the present numerical
experiments. Therefore, we do not make any broader claim regarding a universal
scaling law from these data alone. Nevertheless, the trend provides a useful
indication of how the complexity of the adaptive Gaussian representation grows
for these multidimensional Morse calculations.

\subsection{Double Well Potentials} \label{sec:double-well} As a next
class of more complicated examples, we demonstrate VAGD method on the
double-well potential. These systems are dynamically strongly
correlated in time, in the sense that no single classical trajectory
can capture the true dynamics. The deep tunneling observed in these
systems cannot be described by simple single trajectory methods, and
therefore, single Gaussian ansatz fails. Time-sliced TGA has been
shown to be able to account for the tunneling
event~\cite{kongTimeSlicedThawedGaussian2016}. We demonstrate how VAGD
is able to optimally represent the wave-function in one- and two-
dimensions.

\subsubsection{One-Dimensional Problem}
\begin{figure}
    \centering
    \includegraphics{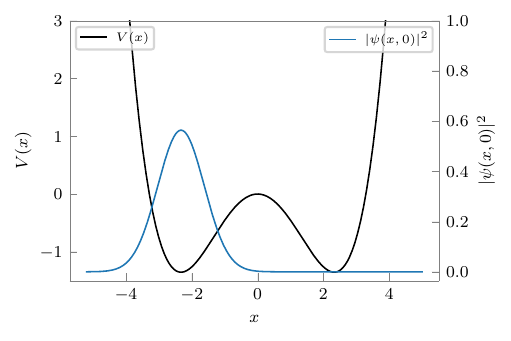}
    \caption{Potential and initial state for the one-dimensional double-well.}
    \label{fig:1ddblwell-setup}
\end{figure}

Following Refs.~\citenum{burantRealTimePath2002, kongTimeSlicedThawedGaussian2016}, the 1D double well
potential is defined to be:
\begin{align}
    V_{\eta}(x) &= \frac{x^{4}}{16\eta} - \frac{x^{2}}{2}, \qquad \eta>0.\label{eq:quartic-double-well}
\end{align}
There are two minima at $x_\pm = \pm 2\sqrt{\eta}$ separated by a barrier at
$x=0$. Consistent with the previous literature, we use
$\eta=1.3544$, so that the minima are located at $x_\pm \approx \pm2.329$. We use an initial GWP localized in the left well,
\begin{align}
    \Psi(x, 0) &= \pi^{-1/4}\exp\left(-\frac{1}{2}\left(x+2\sqrt{\eta}\right)^2\right),
\end{align}
with zero momentum. The potential and the probability
distribution corresponding to $\Psi(x, 0)$ are shown in
Fig.~\ref{fig:1ddblwell-setup}. The initial wave packet is completely localized
in the left well with almost no penetration into the right well. In fact there
is negligible initial population on $x>0$, with the probability of finding the
particle at $x>0$ being $4.97\times 10^{-4}$. This setup can therefore be
treated as a case of deep tunneling. It is also known that in absence of
time-slicing standard semiclassical techniques like Herman-Kluk cannot capture
the tunneling physics~\cite{burantRealTimePath2002}. Though some higher order or
variational single GWP dynamics methods~\cite{faouPoissonIntegratorGaussian2006,
moghaddasifereidaniHighorderGeometricIntegrators2023,
moghaddasifereidaniHighorderGeometricIntegrators2024} may be able to
qualitatively describe deep tunneling phenomena in some cases, in this
particular case of deep tunneling, the variational Gaussian
dynamics~\cite{moghaddasifereidaniHighorderGeometricIntegrators2023} fails. In
contrast, both time-sliced Herman-Kluk~\cite{burantRealTimePath2002} and
TGA~\cite{kongTimeSlicedThawedGaussian2016} can capture the physics in a
numerically exact manner.

For the one-dimensional double-well calculations, the TGA trajectories were
propagated with a time step of $\Delta t=0.01\unit{a.u.}$. The VAGD
reconstruction was carried out every 20 time steps, corresponding to
$N_\text{seg}=20$, using $F_\text{thresh}=0.9999$ and $W_\text{min}=7.5$. The
SOFT reference calculation used the same time step on a uniform grid
$x\in[-8,8]$ with spacing $\Delta x=0.025\unit{a.u.}$.

\begin{figure}
  \centering
  \subfloat[Overlap of TGA and VAGD-TG wave functions with SOFT as a function of time\label{fig:1ddbwell-overlap}]{\includegraphics{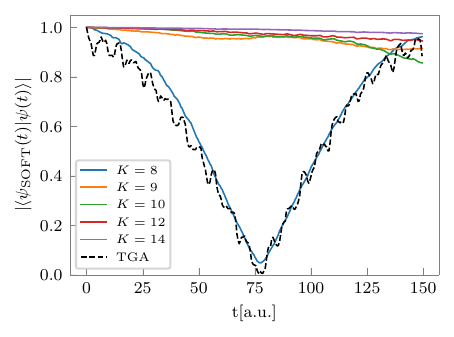}}

  \subfloat[Comparison of $\abs{C(t)}$ at different levels of approximation.\label{fig:1ddbwell-mirror}]{\includegraphics{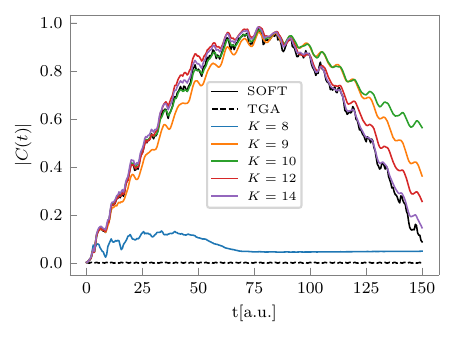}}
  \caption{Accuracy of the VAGD-TG scheme for the 1D double well.}
  \label{fig:1ddbwell}
\end{figure}

\begin{figure}
    \centering
    \includegraphics{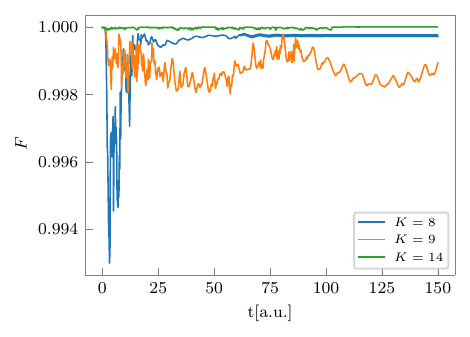}
    \caption{Fidelity of fitting at different stages of dynamics.}
    \label{fig:1ddbwell-fidelity}
\end{figure}

\begin{figure*}
    \centering
    \includegraphics{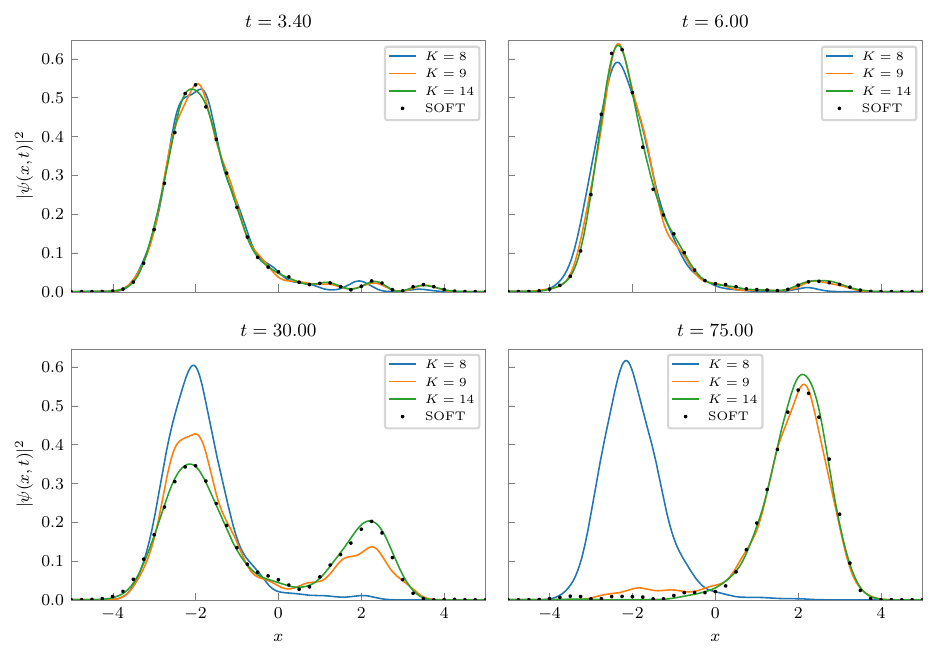}
    \caption{Density snapshots, $\abs{\psi(x,t)}^2$, for the 1D double-well
    system at representative times. The SOFT reference is shown using black
    markers, while VAGD-TG results are shown for $K=8$, $K=9$, and $K=14$.}
    \label{fig:dbwellk8914}
\end{figure*}

\begin{figure}
  \centering
  \includegraphics{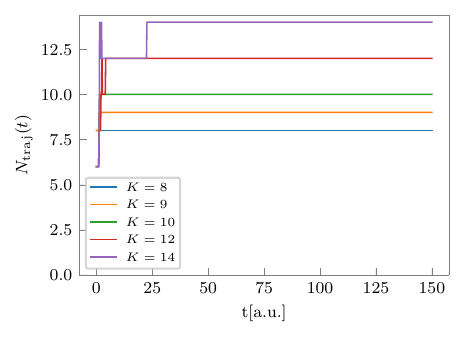}
  \caption{Number of trajectories for the 1D double well system.}
  \label{fig:1ddbwell-traj}
\end{figure}

The shorter segment length, compared with the Morse examples, is
required because the local harmonic approximation deteriorates more
rapidly in the strongly anharmonic double-well potential. More frequent
reconstruction is therefore needed to prevent the semiclassical error
from accumulating between decomposition steps.

In addition to the overlap with SOFT results, following
Ref.~\citenum{kongTimeSlicedThawedGaussian2016}, we correlate the mirror-image
of the initial wave function with the time-propagated one,
\begin{align}
    C(t) &= \int \mathrm{d}x \, \Psi^*(-x, 0) \Psi(x, t),\label{eq:mirror-corr}
\end{align}
measuring the extent of tunneling captured by the various levels of
approximation.

In Fig.~\ref{fig:1ddbwell}, we show the overlap of the semiclassical solutions
with the SOFT wave function (Fig.~\ref{fig:1ddbwell-overlap}) and the absolute value
of the mirror correlation function, Eq.~\ref{eq:mirror-corr}
(Fig.~\ref{fig:1ddbwell-mirror}). The VAGD procedure used $W_\text{min} = 7.5$.
From the overlaps, we see that the TGA result falters immediately, giving an
overlap of almost 0.0 at $t\approx 75\,a. u.$, at which time the \added{SOFT} wavefunction
has completely tunneled from the left to the right well. This first tunneling
event corresponds to the maximum of $\abs{C(t)}$ at around $t\approx 75\,a. u$.
The subsequent lowering of the correlation function corresponds to the wave
function tunneling back to the initial well.

Notice in Fig.~\ref{fig:1ddbwell-mirror}, the TGA wave function does not tunnel at
all. This is expected because of its purely classical nature. It is interesting
that in our VAGD-TG simulations, upto $K=8$, the wave function does not really
tunnel. As we move from $K=8$ to $K=9$, there is a jump in the amount of
tunneling captured. Qualitatively, all the essential features of $\abs{C(t)}$ are
captured by a $K=9$ calculation, though the quantitative agreement is still
missing. It is the second tunneling event back to the original well that proves
to be even more challenging. Using a maximum of $K=14$ trajectories, we start to
get quantitative agreement of the correlation function upto a maximum time of
$t=150$. This is a significant improvement over the original TSTG
method~\cite{kongTimeSlicedThawedGaussian2016}, which used 2048 trajectories
(without filtering), and the TS Herman-Kluk
method~\cite{burantRealTimePath2002}, which used around 525 trajectories.
VAGD-TG, in comparison, requires tens of trajectories to get the numerically
exact result.

To explore the dramatic improvement observed between 8 and 9
trajectories, we try to understand how well the decomposition is happening. The
fidelity achieved by the neural network tells us the accuracy of the output wave
function as a representation of the input wave function. We plot this as a
function of time in Fig.~\ref{fig:1ddbwell-fidelity}. With 8 trajectories, we
see that the wave function has a poor representation right from the first few
steps of decomposition. Any error during the fitting procedure would have a
cumulative effect in that the dynamics of the approximate wave function will
over time deviate from the true result. So the relatively low fidelity of the
8-trajectory run along with the sustained period of this low fidelity, means
that the physics is not going to be faithfully captured. We see that for the
9-trajectory run, the fidelities achieved are higher, leading to a better
approximation of the tunneling dynamics. For the quantitatively correct case of
14-trajectories, the fidelities are practically unity throughout the run.
Consequently there is no drift from the true result. The physical intuition is
that the exponentially decaying tail, which governs tunneling dynamics, is not
captured properly at low numbers of GWPs. Once enough GWPs are used to
faithfully represent the tail, the tunneling is also reproduced properly. (It is
also noticed that when the high target fidelities are not achieved, there is a
much greater variance in the output dynamics with the starting guess of the
neural network parameters. This variability goes away when the number of
Gaussian wavepackets is large enough to give a faithful representation of the
input wave function. This is demonstrated in the supplementary information.)

\added{A different perspective on this qualitative change can be obtained by
looking at the agreement of the SOFT and the VAGD-TG wave functions at various
time points. Figure~\ref{fig:dbwellk8914} shows this data for $K=8,9,$ and 14.
Even at a time as short as $t=3.40$ or 6.00, we already see that $K=8$ is unable
to describe the smaller peaks along the right tail with proper accuracy. Now, 
tunneling is mediated through these tail features that extend into the
classically forbidden regions. Consequently the ability to describe this is
crucial for capturing deep tunneling. As we proceed to longer times, we see this
and later representational error in $K=8$ grow to limit its ability to show
tunneling. $K=9$ qualitatively gets all the features, while at $K=14$,
consistent with the overlaps, the dynamics is nearly quantitatively
reproduced. The real and imaginary parts are also separately shown in the
supporting information (Figures 7--9).}\comment{Wave function decompositions for
1D tunneling $K=8$ vs $K=9$.}

In Fig.~\ref{fig:1ddbwell-traj}, we show the actual number of
trajectories used as a function of time for different values of
$K$. We see a difference from the previous examples. Here, the number of
trajectories almost immediately jumps to the maximum value owing to the
significantly greater degree of anharmonicity. Even the adaptive algorithm is
unable to effectively restrict the $N_\text{traj}(t)$ to extremely low values.
If we had not capped it, the number would have been even larger. However, the
perfect overlaps tell us that the simulations are converged with respect to $K$.
Using larger values of $K$ would have resulted in the adaptive algorithm making
the $N_\text{traj}(t)$ significantly higher than what is required.

\subsubsection{Tunneling in a Two-Dimensional Double-Well} \label{sec:double-well-2d}\comment{Two stage optimization used leading to substantial improvement of dynamics at much lower numbers of trajectories.}
Finally, we explore the 2D double well studied by Kong \textit{et
al.}~\cite{kongTimeSlicedThawedGaussian2016} The model potential couples two
strongly correlated degrees of freedom and is given by
\begin{align}
    V_\eta(x_1, x_2) = \frac{x_1^4}{16\eta} - \frac{x_1^2}{2} + \frac{x_2^2}{2} + \frac{x_1x_2}{2}, \qquad\eta>0.
\end{align}
As in the one-dimensional case, the quartic term generates a
double-well structure along $x_1$, while the bilinear coupling term
$(x_1x_2/2)$ induces a strong correlation between the coordinates. The
initial state is chosen as a single normalized GWP localized in the
left well along $x_1$,
\begin{align}
    \Psi(\vec{x}, 0) &= \pi^{-1/2}\exp\left(-\frac{1}{2}\left((x_1 + 2\sqrt{\eta})^2 + x_2^2\right)\right).
\end{align}     

\begin{figure}
  \centering
  \includegraphics{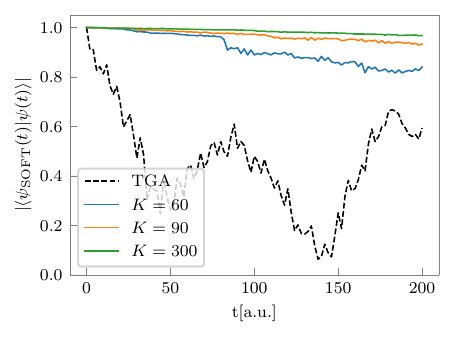}
  \caption{Overlap of VAGD-TG and TGA method for 2D double well potential with SOFT.}
  \label{fig:2d_doublewll}
\end{figure}

\begin{figure}
  \centering
  \includegraphics{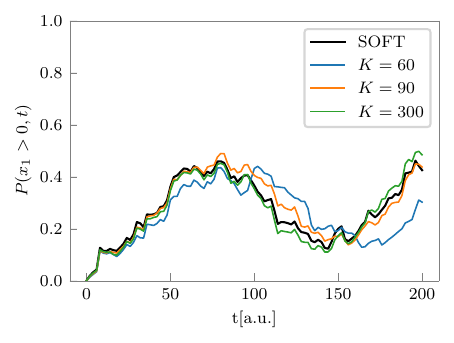}
  \caption{Evolution of tunneling population with time for 2D double well.}
  \label{fig:2d_doublewll_tunneling}
\end{figure}

Three VAGD-TG simulations were run with $\Delta t = 0.01\unit{a.u.}$, and
Fig.~\ref{fig:2d_doublewll} compares their results with TGA. The numerically
exact reference calculations were performed with the SOFT method on a uniform
grid with $x_1, x_2 \in [-15, 15]$ (spacing $\Delta x = 0.1\unit{a.u.}$) and
time step $\Delta t = 0.001\unit{a.u.}$. (Convergence is achieved even at
$\Delta t = 0.01a.u.$) \added{For the present set of parameters, the $K=60$ and $K=90$
calculations were carried out with $W_\text{min}=10$ along both dimensions and
$N_\text{seg}=10$, while the improved $K=300$ calculation used $W_\text{min} =
15$ and $N_\text{seg}=10$. All calculations were run at
$F_\text{thresh}=0.99995$. The $W_\text{min}$ of 15 enforces narrower, more
localized Gaussians, so that the local harmonic approximation remains more
reliable over each segment. Since this is a more difficult optimization problem,
after reaching a minimal fidelity using the neural network, a first-order direct
optimization of the GWPs was used to hit the target fidelity.}\comment{New, better results for 2D double well.}

Figure~\ref{fig:2d_doublewll} compares the overlap of the simulations using
different number of GWPs with the SOFT result. As expected, TGA fails rapidly
for this strongly correlated anharmonic tunneling problem. Its overlap with the
exact SOFT result drops quickly at early times and then exhibits large
oscillations, indicating that a single thawed Gaussian is unable to represent
the evolving correlated structure of the wave function. In contrast, all VAGD-TG
calculations retain substantially higher overlap throughout the propagation
window. \added{The $K=60$ result already captures a large part of the
wave-function structure, but shows a systematic loss of overlap at later times
especially beyond $t\approx75\unit{a.u.}$. Increasing the GWP expansion
to $K=90$ gives a clear improvement and keeps the overlap high over the full
propagation window, implying that the bulk of the physics has been completely
accounted for. The $K=300$ calculation is nearly converged on the scale of the
plot, remaining closest to the SOFT reference throughout the dynamics.}

To examine this more directly, in
Fig.~\ref{fig:2d_doublewll_tunneling} we show the population that has
tunneled to the right half-plane,
\begin{align}
    P(x_1>0,t) = \int_{0}^{\infty}\mathrm{d}x_1\int_{-\infty}^{\infty}\mathrm{d}x_2\,
    |\Psi(x_1,x_2,t)|^2.
\end{align}
\added{The $K=60$ calculation captures the onset of tunneling and the
population transfer to the right half-plane. The recurrence is also
qualitatively captured. However, it shows visible deviations from SOFT in terms
of the timescales involved. With $K=60$, the entire tunneling dynamics seems to
have slowed down.}

\added{The VAGD-TG results for $K=90$ and $K=300$ track the SOFT calculation
rather well throughout the entire simulation. The rise of the
population during the first tunneling event, its subsequent decrease
as the wave packet returns, and the re-emergence of the population at
later times are all captured correctly. Despite the drop in the overlap
of the $K=90$ run, it gets the correct physical picture for the tunneling
dynamics. Pushing $K$ to 300 brings quantitative agreement with the SOFT results
for both the overlap and the tunneling population. (The evolution of the
positions are shown in the supporting information, and they follow a similar
trend to the tunneling population.) In comparison, the TSTG calculation of the
same system required more than a million
trajectories~\cite{kongTimeSlicedThawedGaussian2016} in comparison to the
current VAGD-TG simulation using around a hundred GWPs. A final comment is due
about the flexibility of the GWP representations. In these simulations, we have
put the same $W_\text{min}$ along both $x_1$ and $x_2$. Given that the
anharmonicities intrinsic to the two dimensions are different, it is expected
that the $x_1$ direction would possibly require thinner wave functions for the
local harmonic approximation to be valid, while $x_2$ may be able to accommodate
wider wavepackets. Such ideas may be leveraged in the future for decreasing the
number of GWPs and increasing the accuracy of the dynamics.} 

\section{Conclusion}\label{sec:conclusion}

The time-slicing approach~\cite{hellerWavepacketPathIntegral1975,
burantRealTimePath2002, kongTimeSlicedThawedGaussian2016} enables semiclassical
simulations to achieve full quantum accuracy at the cost of propagating
additional classical trajectories. The primary challenge, however, lies in
decomposing the multidimensional time-evolved wave function into a basis of
Gaussian wave packets. Monte Carlo–based approaches might appear to offer a
natural solution to this problem, but in practice they are fundamentally limited
by the dynamical sign problem~\cite{makri1988MonteCarloPathIntegration,makri1987MonteCarloOscillatoryIntegrands,mak1990SolvingSignProblem}. Even when Monte Carlo sampling can be converged,
the number of trajectories required becomes prohibitively large, with the
statistical error scaling as $N^{-1/2}$ for $N$ samples in the absence of a sign
problem. Through our Variational Adaptive Gaussian Decomposition (VAGD) scheme,
we demonstrate that the essential physics encoded in the wave function can
instead be captured using a relatively small number of carefully selected
trajectories, providing a scalable and quadrature-free route to time-sliced
thawed Gaussian dynamics.

VAGD provides a variationally optimal decomposition of a given wave function in
terms of GWPs. Its main advantage is that the effective ``basis'' is optimized
specifically for the wave function under consideration through an
autoencoder–decoder architecture. This decomposition scheme can be utilized in a
variety of time-sliced architectures built on semiclassical or quantum schemes.
Most prominently, one can think of combinations with time-sliced versions of
Herman-Kluk semiclassical~\cite{burantRealTimePath2002}, or thawed Gaussian
approximation~\cite{kongTimeSlicedThawedGaussian2016} or even with Rothe's
method of
propagation~\cite{rotheZweidimensionaleParabolischeRandwertaufgaben1930,
schraderTimeEvolutionOptimization2024,
schraderMultidimensionalQuantumDynamics2025}. In the present work, we
demonstrate the combination with time-sliced thawed Gaussian dynamics (VAGD–TG).
The method has been tested on the Morse potential and on one- and
two-dimensional tunneling problems. Dimensional scaling was further examined
using independent Morse oscillators up to four dimensions, where the
computational cost was found to grow only mildly with dimensionality. Our
results suggest a polynomial scaling behavior, although the precise performance
can not yet be commented on from the numerical explorations. \added{In fact, the
scaling is dependent on the correlation between the dimensions and the
overall anharmonicity of the system under study.} Nevertheless, VAGD promises a
substantial improvement over the exponential scaling typically encountered in
conventional approaches. \added{Additionally, it provides a systematic route to
converging to the fully quantum result, giving us an observable-specific handle
on accuracy. One can decide upon the trade-off of increasing the number of
trajectories to increase the accuracy for the specific problem at hand.}

Particularly encouraging is the performance observed in tunneling problems. In
the one-dimensional case, the tunneling dynamics becomes visible with as few as
10 trajectories, while quantum accuracy is achieved with only 14 trajectories,
representing an improvement of roughly two orders of magnitude compared with the
$\sim 10^3$ trajectories required by TSTG. In two-dimensional tunneling,
similarly accurate results are obtained with approximately 300
trajectories, whereas TSTG required millions. VAGD-TG is extremely effective in
these tunneling scenarios, where the wave function acquires interference-driven
non-trivial spatial structure which cannot be captured by a single Gaussian. The
adaptive multi-Gaussian representation that VAGD provides thus yields an
optimally adequate representation of the complexity.

Beyond its practical efficiency, VAGD additionally offers an appealing
conceptual perspective on semiclassical dynamics. The number of Gaussian wave
packets required in the decomposition naturally reflects the degree of
non-classical structure present in the evolving wave function. Systems
exhibiting largely classical behavior can be represented with only a small
number of trajectories, whereas strongly quantum phenomena, such as interference
and tunneling, require richer Gaussian representations. In this sense, the
adaptive Gaussian basis produced by VAGD provides a direct measure of the
intrinsic quantum complexity of the dynamics. \added{Future work will probe
applications of VAGD to study the quantum dynamics of many-particle systems.
Additionally, developments will focus on simultaneously decreasing the
decomposition time and the number of trajectories even further making it
suitable for use in combination with \textit{ab initio} applications. Coupling
with other dynamical methods like VGD or single-Hessian TGA would also be
evaluated. Going forward VAGD promises to provide a platform for developing and
refining methods aimed at scalable simulations of large dimensional systems that
systematically converge to the full quantum results.}


\section{Supplementary Material}
\added{The accompanying supplementary material includes a detailed description of the
implementation of the neural network and training protocol, a demonstration of
convergence of spectra versus the overlap for the 1D Morse oscillator, a
discussion on the variance in 1D double well tunneling case, and graphs showing
the placements of the GWPs in this case, dynamics of positions for the 2D double
well.}\comment{Updates to SI.}

\appendix
\section{Fidelity Calculations}\label{sec:fidelity}
Optimizing the representation means that the fidelity has to be evaluated
multiple times. The overlap calculation is a multidimensional integral and can
prove to be a bottle-neck for multidimensional problems. However, in practice,
fidelity can be evaluated entirely analytically when both $\ket{\psi}$ and
$\ket{\Psi}$ are expressed as superpositions of GWPs. The overlap that is
required for the fidelity is given as $\braket{\psi}{\Psi} =
\sum_{i=1}^{N_\text{inp}}\sum_{j=1}^{N_\text{out}} c_i^* C_j
\braket{\phi_i}{\Phi_j}$. All the overlaps of the pairs of GWPs have closed form
expressions. For two $d$-dimensional GWPs of the form given in
Eq.~\eqref{eq:gwp}, with $\ket{\phi_i}$ parametrized by
$(\vec{q}_i,\vec{p}_i,A_i,\gamma_i)$ and $\ket{\Phi_j}$ parametrized by
$(\vec{Q}_j,\vec{P}_j,\mathcal{A}_j,\Gamma_j)$, define
\begin{align}
    B_{ij} &= \mathcal{A}_j - A_i^*, \\
    \vec{b}_{ij} &= \vec{P}_j - \vec{p}_i - \mathcal{A}_j\vec{Q}_j + A_i^*\vec{q}_i.
\end{align}
Then the elementary overlap is
\begin{align}
    &\braket{\phi_i}{\Phi_j} = \sqrt{ \frac{(2\pi i\hbar)^d}{\det B_{ij}} }\,e^{\frac{i}{\hbar}(\Gamma_j-\gamma^*_i)-\frac{i}{\hbar} \left(\vec{P}_j^{\,\top}\vec{Q}_j-\vec{p}_i^{\,\top}\vec{q}_i\right)}\nonumber\\
    &\times\exp\left( -\frac{i}{2\hbar} \left[ \vec{b}_{ij}^{\,\top}B_{ij}^{-1}\vec{b}_{ij} - \left( \vec{Q}_j^{\,\top}\mathcal{A}_j\vec{Q}_j - \vec{q}_i^{\,\top}A_i^*\vec{q}_i \right) \right] \right).
    \label{eq:gwp-analytic-overlap}
\end{align}
Thus, the loss defined in Eq.~\eqref{eq:loss} can be efficiently evaluated using
these overlap matrices.

To get the full overlap, we need $N_\text{inp}\times N_\text{out}$
elementary overlap computations, the cost of each of which also grows
with the dimensionality of the problem. So, despite having the exact
analytical forms, the computation can get costly. Because each of the
computation is independent, they can be trivially parallelized.

\added{The fidelity, $F$, is defined as $F =
\frac{\abs{\braket{\psi}{\Psi}}}
{\sqrt{\braket{\psi}{\psi}\braket{\Psi}{\Psi}}}$, where
$\ket{\psi}$ is the candidate output obtained from the optimization procedure,
and $\ket{\Psi}$ is the target state, both represented as sums of Gaussian
wavepackets. The norm of the target state can obviously be cached once per
decomposition step. Additionally, by judiciously precomputing and storing
various portions of the Eq.~\eqref{eq:gwp-analytic-overlap} corresponding to the
known and unchanging target state, one can significantly reduce the
computational burden of the fidelity calculations.}\comment{Comment on optimal
implementations.}

Further optimizations can be brought in, by filtering the overlaps by
distance of the centers of the wavepackets. If the centers of two GWPs
are far apart in comparison to their combined width, then the overlap
will become negligible. One can consequently avoid the matrix
operations for such pairs altogether, thereby improving the speed of
the overlap computation.

\section{Nature of GWPs generated: 1D Morse Example}\label{sec:nature-gwp}\comment{Corrections to the GWP decomposition and more accurate data.}
\begin{figure}
    \centering
    \subfloat[1D Morse oscillator and initial state as a Gaussian wave packet.\label{fig:morse_init}]{\includegraphics{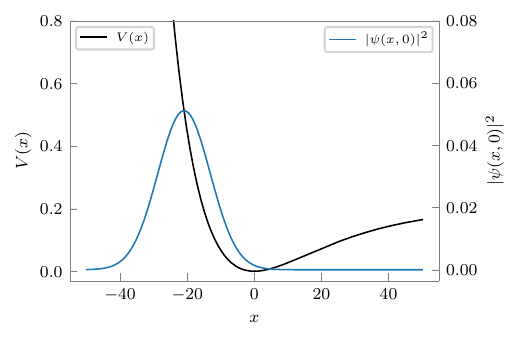}}
    
    \subfloat[Expectation value of position with time.\label{fig:morse_time}]{\includegraphics{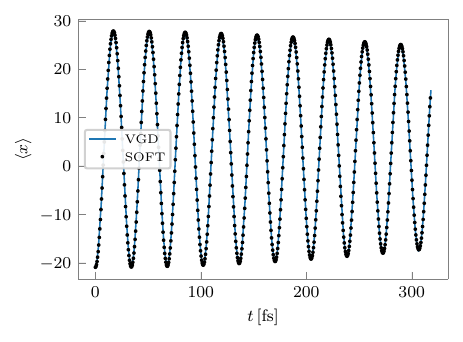}}
    \caption{1D Morse oscillator ($\chi=0.005$), initial condition, and dynamics.}
\end{figure}
\begin{figure*}
    \includegraphics{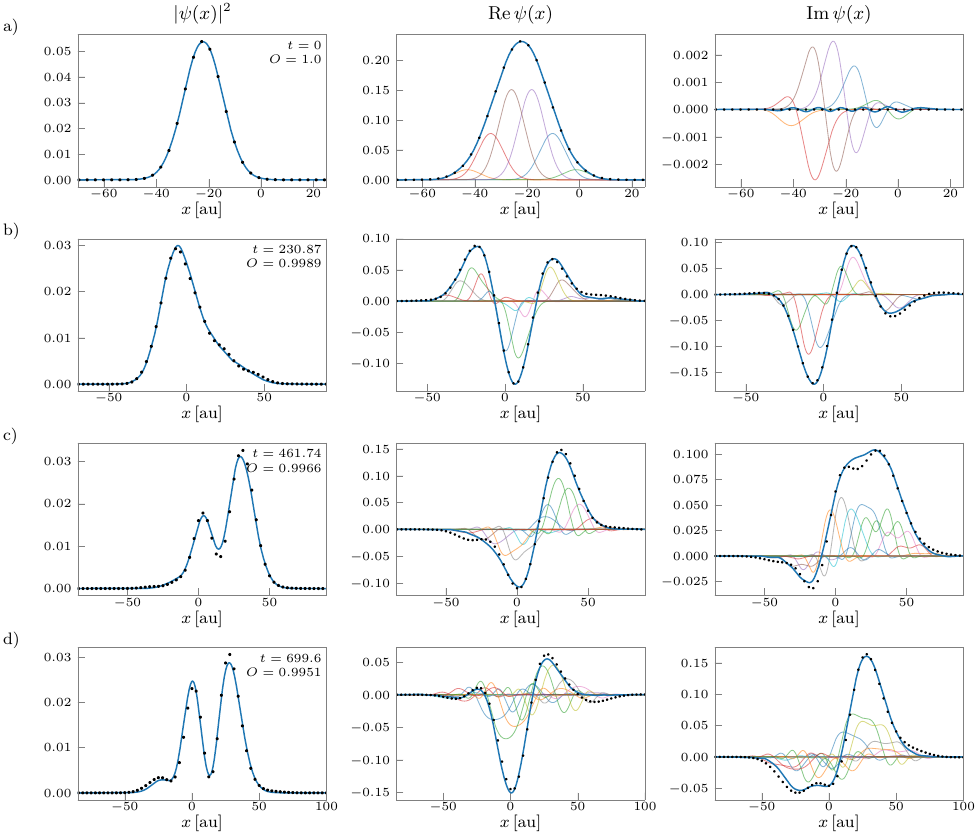}
    \caption{VAGD decomposition at different times. Thin colored lines indicate individual GWPs. Markers: true SOFT wavefunctions; solid blue line: VAGD-TG wavefunction; $O$ is the overlap of SOFT and VAGD results.}\label{fig:vagd-morse}
\end{figure*}
We demonstrate the time-evolution of an initial Gaussian wave packet under a
Morse potential and its VAGD decompositions at different time snapshots. As an
example we use the $\chi=0.005$ case from Sec.~\ref{sec:1Dmorse}.
Figure~\ref{fig:morse_init} shows the potential and the initial wave function.
The evolution of position with time is also shown in
Fig.~\ref{fig:morse_time}. At different decomposition steps \added{along the VAGD-TG trajectory}, VAGD positions
varying numbers of GWPs optimally. In Fig.~\ref{fig:vagd-morse}, we show the decomposition of
the wave function into GWPs at different times. Notice how even the initial time
wave function, despite being a single GWP is decomposed as a superposition of
multiple GWPs owing to the width limit. We see six GWPs being involved in the
decomposition at $t=0$ (Fig.~\ref{fig:vagd-morse}~(a)). at $t=230.87$ in
Fig.~\ref{fig:vagd-morse}~(b), where we see a shoulder start to develop the number of GWPs is increased to 14.
However, by the time we reach the bimodal wave function in
Fig.~\ref{fig:vagd-morse}~(d), 16 GWPs are required for a proper description.
Not only is the number of GWPs dependent on the input wave function, the details
of the individual GWPs change as well from case to case, adapting to the
requirements.

\section{Cost of VAGD-TG}\label{sec:cost}\comment{More data and details of decomposition times added.}

We present a preliminary empirical exploration of the cost of VAGD training for
our particular implementation. It should be noted that the following study is
not an optimized benchmark by any measure. While several algorithmic
improvements can still be incorporated, this provides an intuitive picture into
the performance for an initial implementation. We use the 1D and 2D double well systems to study the run-times. Given the extremely small dimensionality of the problem, the actual time spent on the classical trajectories are negligible in both cases and all the run-time comes from the decompositions. All the timing data reported in this section were obtained on a single NVIDIA V100 GPU.

\begin{figure*}
\centering
\subfloat[1D double well\label{fig:timing_1d_violin}]{\includegraphics{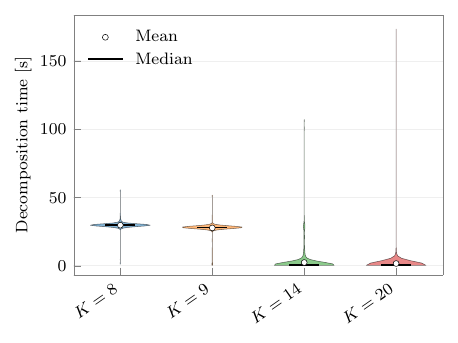}}
~\subfloat[2D double well\label{fig:timing_2d_violin}]{\includegraphics{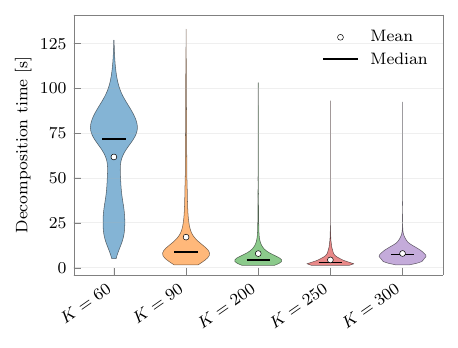}}
\caption{Distribution of decomposition times along a VAGD-TG simulation for the
1D double-well calculation where a time-evolved $K$-GWP input state is decomposed into a $K$-GWP output state reenforcing all the conditions.}
\end{figure*}

We show the distribution of decomposition times along the VAGD-TG simulation of
the one-dimensional double well for changing $K$ in
Fig.~\ref{fig:timing_1d_violin}. There are some interesting features. First, it
might come as a surprise that the $K=8$ simulations spend the maximum time in
the VAGD procedure on average. This in hind-sight makes sense for two reasons:
(a) for this simple problem, neural network optimization is an overkill and the
startup costs are really not worth it; and more importantly, (b) owing to the
low representational power of the $K=8$ \textit{ansatz}, the code repeatedly
fails to achieve the target fidelity threshold and keeps searching for the
optimal representation for several further epochs. This latter issue can easily
be suppressed by changing meta-parameters of the VAGD algorithm. Similar results
are obtained for the K=9 run. Although the \textit{ansatz} is flexible enough to
show the first tunneling event qualitatively it is also not enough to hit the
target fidelity for most of the decomposition steps. Next, notice that once the
representation becomes flexible enough, at $K=14$ and $K=20$, the median time
taken drops down to very low values. One notices that the variance however is
very large. A few decomposition steps take exceptionally long times. The reason
is likely to be the failure of the warm-start \textit{ansatz}.


Several of these features carry over to the 2D case as shown in
Fig.~\ref{fig:timing_2d_violin}. In this case even the $K=60$ run shows the high
decomposition time characteristic of the representation not being sufficiently
flexible. The cost drops down substantially at $K=90$. Interestingly one notices
a not insignificant decrease in the mean and median decomposition times from
$K=90$ to 200 to 250. The increase in representational power substantially aids
the optimization processes in this regime, though the number of calculations
required for the fidelity calculations increase. From $K=250$ to $K=300$, we
notice the increased cost of the fidelity and optimization of larger set of
parameters become dominant leading to an increase of the total decomposition
time.

While this exploration of the decomposition times are preliminary, and several
further optimizations can be implemented to make the algorithm faster, a couple
of conclusions that stand out is that there is only a very weak dependence of
the run-time with the number of GWPs being used in the representation, once
\textit{ansatz} is flexible enough to give a proper description of the wave
function. Going from one- to two-dimensions also does not substantially increase
the cost of a decomposition step. This holds great promise for systems with
larger dimensions and which require numerous GWPs for adequate descriptions.

\bibliography{references}
\end{document}